# Dielectric Optical Cloak

**Jason Valentine[1]\*, Jensen Li[1]\*, Thomas Zentgraf[1]\*, Guy Bartal[1] and Xiang Zhang[1,2]**

[1]*NSF Nano-scale Science and Engineering Center (NSEC), 3112 Etcheverry Hall,*

*University of California, Berkeley, California 94720, USA*

[2]*Material Sciences Division, Lawrence Berkeley National Laboratory, Berkeley, California*

*94720*

*\*These authors contributed equally to this work*

Invisibility or cloaking has captured human's imagination for many years. With the recent advancement of metamaterials, several theoretical proposals show cloaking of objects is possible, however, so far there is a lack of an experimental demonstration at optical frequencies. Here, we report the first experimental realization of a dielectric optical cloak. The cloak is designed using quasi-conformal mapping to conceal an object that is placed under a curved reflecting surface which imitates the reflection of a flat surface. Our cloak consists only of isotropic dielectric materials which enables broadband and low-loss invisibility at a wavelength range of 1400-1800 nm.

For years, cloaking devices with the ability to render objects invisible were the subject of science fiction novels while being unattainable in reality. Nevertheless, recent theories including transformation optics (TO) and conformal mapping [1-4] proposed that cloaking devices are in principle possible, given the availability of the appropriate medium. The advent of metamaterials [5-7] has provided such a medium for which the electromagnetic material properties can be tailored at will, enabling precise control over the spatial variation in the material response (electric permittivity and magnetic permeability). The first experimental demonstration of cloaking was recently achieved at microwave frequencies [8] utilizing metallic rings possessing spatially varied magnetic resonances with extreme permeabilities. Such stringent requirements are normally accompanied by strong dispersion, resulting in a cloak that only works in a narrow frequency range. Additionally, the strong magnetic response obtained in microwave metamaterials [5-7,9], where metal is close to being perfect, cannot be sustained at optical frequencies due to the higher loss and the kinetic inductance of the electrons, prohibiting a simple scaling to optical frequencies [10-12].

Recently, a number of new forms of cloaking have been proposed to mitigate these constraints [13-18], relaxing the requirement of extreme material properties. Of special interest is the so-called carpet cloak. Previous cloak designs either compress the object into a singular point or line. The carpet cloak compresses an object in only one direction into a conducting sheet [16]. When the object is sitting under a curved reflecting surface with the carpet cloak on top of it, the object appears as if it is the original flat reflecting surface, so it is hidden under a 'carpet'. Such a topological change of the cloak avoids both material and geometry singularities [19]. Unlike previous cloaking transformations, it is non-singular and generates a smaller range for the material properties. Quasi-conformal mapping is also employed so that all the originally square cells are transformed to rectangles of a constant aspect ratio. While all other mappings result in a highly anisotropic cloak profile, the quasi-conformal mapping can make the transformed cells almost square, so that the anisotropy of the medium is minimized to a point where it can be neglected. This results in a modest range of isotropic indices for the cloak. The approach allows the use of non-resonant elements (e.g., conventional dielectric materials) and offers the possibility to achieve low-loss and broadband cloaking at optical wavelengths, rendering an object truly undetectable with visible or infrared light. Carpet cloaking was recently realized experimentally at microwave frequencies, utilizing non-resonant metallic elements [20]. However, even with the advances in optical metamaterials [21-23], scaling sub-wavelength metallic elements and placing them in an arbitrarily designed spatial manner still remain challenging at optical frequencies.

In this report, we experimentally demonstrate optical cloaking utilizing a dielectric carpet cloak design. This form of cloaking is also shown to be not only isotropic, but also low-loss and broadband. The invisibility is demonstrated within a Silicon (Si) slab waveguide where the cloak region is obtained by varying the effective index of refraction in a two-dimensional (2D) space. This index profile is designed using quasi-conformal mapping and realized by fabricating a 2D sub-wavelength hole lattice with varying density. The cloak is placed around a reflecting curved surface (bump) and consequently, a light beam incident on the bump exhibits a reflection profile identical to that of a beam reflected from a flat surface. Therefore, any arbitrarily shaped object placed behind the bump will maintain the reflectance of a smooth, flat surface, rendering the object invisible. This approach represents a major step towards general transformation optics [24-26] which has so far remained a challenging endeavor due to increased metal loss and fabrication limitations at optical frequencies, with only basic optical applications (e.g., lenses) brought to reality [27-29]. It simplifies the realization of an arbitrary 2D sub-wavelength effective index profile, by using the simple and uniform geometry of a hole array with variable density. This allows for easy fabrication and scaling, opening the possibility for a large variety of TO devices at visible and infrared wavelengths. This methodology eliminates the need for metallic elements and anisotropy with spatial variations at a deep sub-wavelength scale and a size of tens of nanometers.

The carpet cloak metamaterial was fabricated on a silicon-on-insulator (SOI) wafer consisting of a 250-nm-thick silicon layer separated by a 3-µm-thick silicon oxide ($SiO_2$) slab from a Si wafer substrate (Fig. 1a). The 250 nm Si layer serves as an optical slab waveguide where the light is confined in vertical dimension and can freely propagate in the other two dimensions. While this report will focus on the transverse magnetic waveguide mode (TM-mode) for the experimental demonstration of the cloaking, it is also possible to design the cloak for the TE waveguide mode. The reflecting surface was fabricated by using focused ion beam milling (FIB) to etch through the Si layer and partially through the $SiO_2$ slab near the edge of the SOI substrate, making the surface accessible for directional deposition of metal from the side.

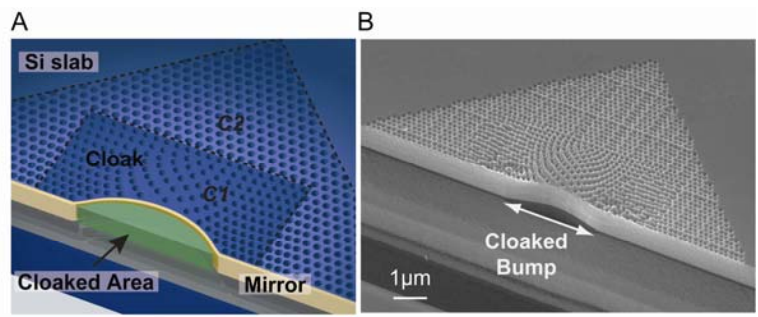

**Fig. 1.** The carpet cloak design that transforms a mirror with a bump into a virtually flat mirror. (**A**) Schematic of fabricated carpet cloak showing the different regions where *C1* is the gradient index cloak and *C2* is a uniform index background. The cloak is fabricated in a SOI wafer where the Si slab serves as a 2D waveguide. The cloaked region (marked with green) resides below the reflecting bump (carpet) and can conceal any arbitrary object. The cloak will transform the shape of the bump back into a virtually flat object. (**B**) SEM image of fabricated carpet cloak. The width and depth of the cloaked bump are 3.8 µm and 400 nm, respectively.

The carpet cloak region of the device (Fig 1a) is composed of a triangular region (marked as C2) with uniform hole pattern, serving as a background medium with a constant effective index of 1.58, and a rectangular cloak region (marked as C1) with the two dimensional variable index profile. The hole pattern was created by milling holes of constant diameter (110 nm) through the Si layer with FIB. The holes were milled with varying densities yielding the desired spatial index profile seen in Fig. 1b. Additionally, two gratings were fabricated for coupling the light into and out of the Si-waveguide. Directional deposition of 100 nm gold (Au) was then performed using electron beam evaporation to create the reflecting surface. An overall image of the device layout can be seen in Fig. 2.

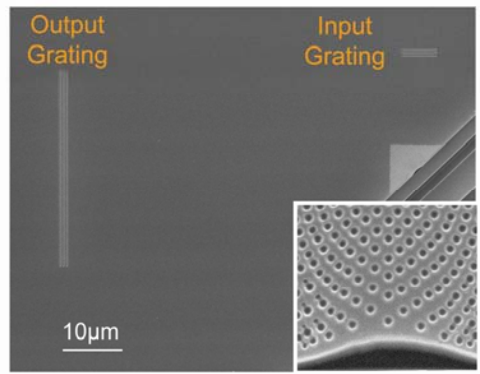

**Fig. 2.** Scanning electron microscope image of carpet cloak layout. Light is coupled into the Si slab waveguide via the input grating which has a width of 7 µm and distance from the cloak of 15 µm. After being reflected at the cloaked surface, the beam profile is detected via the output grating which has a width of 35 µm and a distance from the cloak of 55 µm. The inset shows the central region of the cloak. The hole diameter is 110 nm.

To unambiguously prove the carpet cloak, we compare the profile of a Gaussian beam reflected from a cloaked bump to that of a similar beam reflected from (1) a flat surface without a cloak and (2) a surface with bump but without a cloak. The tunable light from a femtosecond synchronously pumped optical parametric oscillator (Spectra-Physics, OPAL) was focused at the input grating to launch the fundamental TM wave in the Si slab. A charged coupled device (CCD) camera was then used to measure the light coupled out of the waveguide at the output grating position.

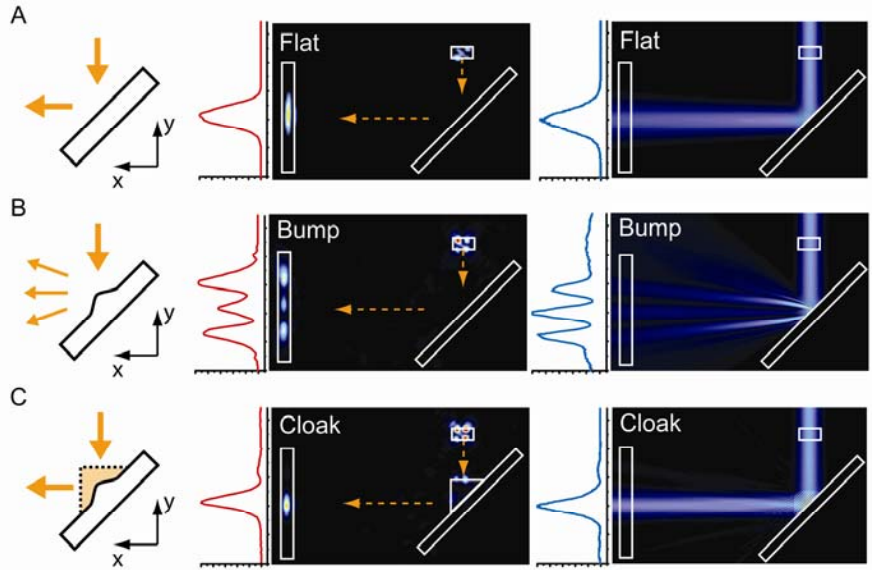

**Fig. 3.** Optical carpet cloaking at a wavelength of 1540 nm. The results for a Gaussian beam reflected from (**A**) a flat surface (**B**) a curved (without cloak) surface, and (**C**) the same curved reflecting surface with cloak. The left panel shows the schematics. The middle panel shows the optical microscope images and normalized intensity along the output grating position. The curved surface scatters the incident beam into three separate lobes while the cloaked curved surface maintains the original profile, similar to reflection from a flat surface. The experimental intensity profile agrees well with the intensity profile ($|E_z|^2$) from 2D simulations (right panel).

Figure 3 displays the images for the three configurations as well as the intensity profile at the output grating. As seen in Fig. 3a and 3b, there is considerable contrast between the reflection from the flat and curved surface. The light reflected from the uncloaked bump shows three distinct spots at the output grating due to the scattering of the bump, which is significantly larger than the wavelength. The flat surface displays the expected Gaussian beam profile, similar to that of the incident wave. To hide the bump on the surface, the designed cloak pattern was placed around the bump in Fig. 3c. Subsequently, the beam profile at the output grating resembles a single reflected beam as is seen with the flat reflecting surface. This demonstrates that the cloak has successfully transformed the curved surface into a flat surface, giving the observer the impression that the beam was reflected from a flat surface. Due to the fact that there is no penetration of light into the bump (through the metal layer), any object could be placed behind it and effectively hidden, making the object invisible. Both 3D and 2D simulations where performed with a commercial finite element method package (COMSOL) to verify the carpet cloak performance. The 2D simulation results for the three configurations are shown in Fig. 3 (3D simulations which show good agreement with the 2D simulations). The simulations show the magnitude of the electric field component in z-direction and show a good agreement with the experimental results. We note that the demonstration of the cloaking effect by using an isotropic profile here is in fact closely connected to the optical conformal mapping in Ref 3 but the phase information is also preserved in our case. To show the angular performance of the cloak, samples were fabricated where the input grating was placed at 30° and 60° with respect to the reflective surface. In both cases, the cloaked bump produced a singular beam at the output as is seen with a flat mirror.

Since the carpet cloak reported here does not rely on resonant elements, it is expected to be nearly lossless and broad-band. Nevertheless, the current nanofabrication technology has a practical limit on the deep sub-wavelength hole size due to the waveguide thickness, causing scattering. Further losses are encountered due to possible residual Gallium left from the FIB. Consequently, the transmission of the carpet cloaking sample was found experimentally to be 58% at a wavelength of 1540 nm. Nonetheless, we emphasize that this reduced transmission is due to experimental imperfections associated with the technique of drilling the holes (FIB) and is not inherent to this cloaking design methodology. For example, electron beam lithography could be used in conjunction with reactive ion etching to completely eliminate the loss associated with Gallium as well as provide for much larger cloaking devices. In this case, almost perfect transmission should be attainable due to low intrinsic loss of silicon or other dielectrics.

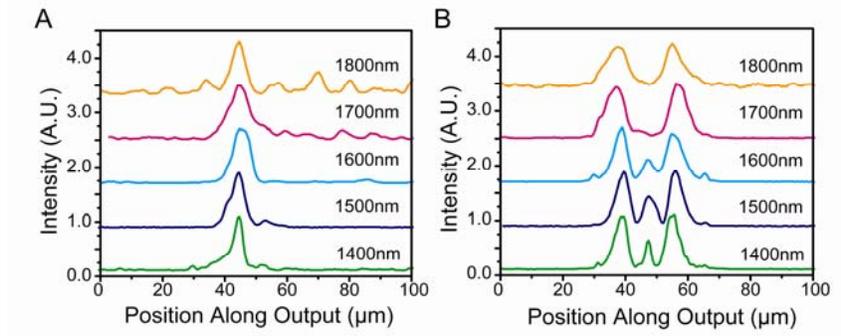

**Fig. 4.** Wavelength dependence of the carpet cloak. Plotted is the intensity along the output grating for a curved reflecting surface (**A**) with cloak and (**B**) without the cloak. The cloak demonstrates broadband performance at 1400 nm – 1800 nm wavelengths. Distinct splitting of the incident beam is observed from the uncloaked curved surface due to the strong scattering of the original beam.

To obtain the bandwidth of the carpet cloak, measurements were performed over a wide range of wavelengths. The output intensity profiles for these measurements are plotted in Fig 4a. For a broad wavelength range from 1400 nm to 1800 nm the beam profile shows a single peak at the output grating, i.e. the cloak performance is largely unaffected by the wavelength change. At wavelengths below 1400 nm the cloak performance suffers due to the fact that the wavelength in the slab waveguide ($\lambda_0/n_{Si}$) becomes comparable to the hole diameter of the cloak pattern, causing increased scattering and breakdown of the effective medium approximation. The effectiveness of the optical cloak can be improved at wavelengths below 1400 nm by using smaller hole diameter and separation, reducing scattering and extending effective medium approximation to shorter wavelength. Unlike the cloak, the bump alone displays a multi-peak output beam (Fig. 4b) which is clearly observed for all wavelengths, indicating the strong perturbation of the beam. The upper wavelength limit for the cloak is ultimately restricted by waveguide cutoff. However, measurements above 1800 nm were not possible due to the CCD sensitivity cutoff.

The experimental demonstration of cloaking at optical frequencies suggests invisibility devices are indeed within reach. The all-dielectric design is isotropic and non-resonance based, therefore promising a new class of broadband and low-loss optical cloaks. It should be noted, that this methodology can also be extended into an air background by incorporating non-resonant metallic elements to achieve indices smaller than one. Furthermore, the quasi-conformal mapping design and fabrication methodology presented here may open new realms of transformation optics beyond cloaking.